# MACHINE DECISIONS AND HUMAN CONSEQUENCES[*]

Teresa Scantamburlo, Andrew Charlesworth, Nello Cristianini

University of Bristol

## ABSTRACT

As we increasingly delegate decision-making to algorithms, whether directly or indirectly, important questions emerge in circumstances where those decisions have direct consequences for individual rights and personal opportunities, as well as for the collective good. This is most evident in the case of certain forms of decision-making in the criminal justice system concerning personal liberty, such as custody, sentencing and parole determinations, but also applies to access to education, housing, social welfare, credit, and insurance. A key problem for policymakers is that the social implications of these new methods can only be grasped if there is an adequate comprehension of their general technical underpinnings.

The discussion here focuses primarily on the case of enforcement decisions in the criminal justice system, but draws on similar situations emerging from other algorithms utilised in controlling access to opportunities, to explain how machine learning works and, as a result, how decisions are made by modern intelligent algorithms or 'classifiers'. It examines the key aspects of the performance of classifiers, including how classifiers learn, the fact that they operate on the basis of correlation rather than causation, and that the term 'bias' in machine learning has a different meaning to common usage.

An example of a real world 'classifier', the Harm Assessment Risk Tool (HART), is examined, through identification of its technical features: the classification method, the training data and the test data, the features and the labels, validation and performance measures. Four normative benchmarks are then considered by reference to HART: (a) prediction accuracy (b) fairness and equality before the law (c) transparency and accountability (d) informational privacy and freedom of expression, in order to demonstrate how its technical features have important normative dimensions that bear directly on the extent to which the system can be regarded as a viable and legitimate support for, or even alternative to, existing human decision-makers.

It is contended that systems which utilise decision-making (or decision-supporting) algorithms, and have the potential to detrimentally affect individual or collective human rights, deserve significantly greater regulatory scrutiny than those systems that use decision-making algorithms to process objects. This may seem a truism, yet it appears that the continuing expansion of machine decision-making in the public and private sectors, often justified on the grounds of efficiency and cost-saving, has yet to attract the degree of informed regulatory analysis and evaluation that its problematic merit.

**INTRODUCTION**

The recent 'machine learning' data revolution has not just resulted in game-playing programs like AlphaGo, or personal assistants like Siri, but in a multitude of applications aimed at improving efficiency in different organizations by automating decision-making processes. These applications (either proposed or deployed) may result in decisions being taken by algorithms about individuals which directly affect access to key societal opportunities, such as education, jobs, insurance and credit. For, example, job applications are often screened by intelligent software[1] for shortlisting, before being seen by human recruiters;[2] college admissions are increasingly informed by predictive analytics,[3] as are mortgage applications,[4] and insurance claims.[5] In those circumstances, individuals are usually voluntarily engaging with the organisation utilising the software, although they may not be aware of its use (Eubanks 2018). However, predicative analytics are increasingly in use in circumstances where the individual's engagement with the organisation is largely or wholly involuntary, e.g. in areas of State intervention such as child safeguarding and access to welfare benefits.[6]

A controversial area of application is that of criminal justice, where use of intelligent software may not directly impact the wider population, but the outcome of the decisions made are of critical importance to both the rights of those subject to the criminal justice system, and to the public's perception of the impartiality and fairness of the process.[7] Software tools are already used in US and UK courts and correctional agencies[8] to support decision-making about releasing an offender before trial (e.g. custody and bail decisions), and to determine the nature and length of the punishment meted out to a defendant (e.g. sentencing and parole). Such software, usually referred to as 'risk assessment tools", have become the focus of public attention and discussion because of claims about their associated risks (Angwin *et al.* 2016 and Wexler 2017) and benefits.[9]

---

[1] A note on terminology: throughout the paper we will use the expressions "intelligent software" and "intelligent algorithms" to refer to the products of AI methods and specifically of machine learning. While we will use the terms "machine decision" and "machine decision-making" to indicate the application of machine learning techniques to decision-making processes.

[2] The human resource software market includes applicant tracking systems, resume screening tools, talent matching and candidate relationship systems. Many companies use proprietary software e.g. *JobVite* (https://www.jobvite.com/) or free applications e.g. *MightyRecruiter* (https://www.mightyrecruiter.com/), which, collect and manage applicants' data, but can also provide functions for parsing and classifying applicants' resumes. Other software such as *Belong* (https://belong.co/) can perform more personalized search by analysing candidates' behaviour on social media platforms like Facebook or Twitter.

[3] E.g. consulting firm EAB (https://www.eab.com) provides technological solutions to enable colleges and universities to achieve the desired target of applicants, track students' activities and offer personalised advises. Institutions using these, or similar tools, include Georgia State University, (https://www.eab.com/technology/student-success-collaborative/ssc-wsj-oct-13), and Saint Louis University, which has developed its own data-driven approach for finding successful students in existing and new geographic markets (Selingo 2017).

[4] E.g. *Experian* (http://www.experian.com), one of the largest UK Credit Reference Agencies, is using machine learning to approve or deny mortgage applications reducing processing time (see e.g. http://www.experian.com/blogs/news/2017/05/25/information-data/). Other lending platforms, like *Upstart* (https://www.upstart.com/) and *SoFi* (https://www.sofi.com) make their assessment also based on applicant's career and education. For a list of companies using machine learning and AI in lending see (Payne 2017)

[5] E.g. SAS (https://www.sas.com/) provides Detection and Investigation for Insurance which uses automated business rules, embedded AI and machine learning methods, text mining, anomaly detection and network link analysis to automatically score claims records. https://www.sas.com/en_us/software/detection-investigation-for-insurance.html

[6] E.g. in the US, Allegheny County was the first jurisdiction to use a predictive-analytics algorithm to identify families most in need of intervention (Hurley 2018). See the Allegheny Family Screening Tool http://www.alleghenycounty.us/Human-Services/News-Events/Accomplishments/Allegheny-Family-Screening-Tool.aspx In the UK "…at least five local authorities have developed or implemented a predictive analytics system for child safeguarding. At least 377,000 people's data has been incorporated into the different predictive systems." (McIntyre & Pegg 2018).

[7] E.g. the case of Eric Loomis in Wisconsin, USA (Liptak, 2017).

[8] For a survey of the tools used in US see Electronic Privacy Information Centre https://epic.org/algorithmic-transparency/crim-justice/). In UK, for example, see the Harm Risk Assessment Tool examined below. Similar tools, such as actuarial methods, have been used in England and Wales since the 1990s and are reviewed in a Ministry of Justice (UK) report (Moore 2015).

In a democratic society, when a human decision-maker, such as a judge, custody officer, or social worker evaluates an individual for the purposes of making a decision about their entitlement to some tangible benefit or burden with real consequences for the relevant individual, it is understood that they should do so in accordance with a commonly understood set of normative principles, for example:

- justice (e.g. equality before the law, due process, impartiality, fairness);
- lawfulness (e.g. compliance with legal rules);
- protection of rights (e.g. freedom of expression, the right to privacy).

It is expected that not only will a decision-maker normally act in conformity with those principles, but also that the exercise of their decision-making powers can, when challenged, be exposed and subjected to meaningful oversight. This requires that the system within which they operate is subject to certain normative obligations, for example, that it is transparent and accountable, and as such facilitates public participation and engagement.

Many decisions will not be subject to specific review or scrutiny processes as a matter of course. Thus, the legitimacy of the system within which they are made rests both upon the reliance we are willing to place on the decision-makers internalising and applying the normative principles – i.e. our having a rational basis on which to develop trust; and the system's effective observance of the normative obligations which provide the capacity and capability to criticise and question – i.e. our having a meaningful ability to exercise distrust. Decision-making systems in which the ability of the public to effectively develop trust or exercise distrust is significantly attenuated are likely to be perceived as illegitimate, and probably dysfunctional.

Ensuring that machine decision-making operates in accordance with normative principles, and takes place within a system that incorporates normative obligations, will be an important challenge, particularly as the precise nature and degree of observance of those principles and obligations may vary according to cultural and disciplinary expectations.[10] The current generation of intelligent algorithms make decisions based on rules learnt from examples, rather than explicit programming, and often these rules are not readily interpretable by humans. There is thus no guarantee that intelligent algorithms will necessarily internalise accurately, or apply effectively, the relevant normative principles; nor that the system within which they are embedded will have the means to facilitate the meaningful exercise of particular normative obligations.

This chapter focuses on the example of criminal justice, because it highlights many of the concerns shared by other applications of machine learning to human decision-making: how are these decisions made? How will this impact our ability to develop trust or exercise distrust effectively? What do we know about the accuracy, fairness and transparency of such decisions? Can biases affect intelligent algorithms, due to their training examples or the conditions in which they are used? What kind of analogies might we use to think about these situations? What technical or legal solutions should be developed?

The principle underlying the use of machine decisions, that a score can be used as an indicator of the risk that a given person will behave in a certain way over a period of time (e.g. she will commit a crime, graduate successfully or fulfil assigned tasks) remains the same for a broad range of uses, even if the consequences of a decision vary significantly in context. Depending on the score produced, the algorithm triggers a determinate response action such as "detaining an offender" or "rejecting a loan application". But a lack of general understanding of how these algorithms work and the impact they may have on people's lives, combined with limited access to information about software and outcomes, renders such decision-making processes opaque, even as an increasing reliance upon algorithms suggests a need for in-depth examination.

---

[9] For a brief examination of the benefits and risks in criminal justice, see the White House Report on Algorithmic Systems, Opportunity, and Civil Rights (2016:19)

[10] E.g. obligations such as transparency and accountability may be more highly valued in some countries, or in decision-making systems, than in others; and understanding of the meaning and/or application of terms such as 'privacy' or 'equality' can differ between lawyers, computer scientists and laypersons.

There is a need for discussion of the individual and social consequences of increased reliance on machine decision-making, and whether systems predicated upon such consequential decisions can support the necessary elements for the development of trust, or exercise of distrust, that are fundamental to the legitimacy of the criminal justice process, or to the legitimacy of administrative decision-making, when humans play a decreasing (or no) role in the process. The requirements for developing trust and exercising distrust in such systems will extend beyond simple guarantees of accuracy, to encompass guarantees that they operate according to key normative principles, such as justice, lawfulness and protection of specific rights. Such limits might include not making use of information that is normatively excluded when reaching a given decision (e.g. the use of 'gender' or 'sexual orientation' in recruitment decisions), respecting some pre-determined notion of fairness, and providing evidence both about how a decision was reached, and that the system was operated within the range of conditions for which it was designed and authorised.

Four benchmarks derived from the normative principles/objectives, which we consider both common to consequential decision-making relating to individuals, and critical to development of appropriate trust relations between stakeholders, are used here to exemplify the claim that a knowledgeable critical assessment of the interplay of key technical and normative concepts is required for the effective and legitimate application of machine decision-making. These benchmarks are accuracy, transparency/accountability, fairness/equal treatment, and rights to privacy/free expression. It is not suggested that these are the only possible or necessary benchmarks, but they are arguably the most representative elements of the public concerns about the encroachment of technology into social structures that permeate the zeitgeist of the 'big data' era. The main objectives here are to demystify the technology behind machine decisions, making it comprehensible and accessible to a non-technical audience, and to suggest how complementary technical and normative perspectives can help us to articulate and interlink specific requirements and remedies.

The following sections describe how a class of algorithms called 'classifiers' work (section 2), examine how they operate in practice in the Harm Assessment Risk Tool (HART), a real-life implementation in the criminal justice sector (section 3) and hypothesise how the four benchmarks might be used to effect and evaluate meaningful trust relations between stakeholders in the context of available legal and technical remedies (section 4).

**MACHINE DECISIONS: THE TECHNOLOGY**

**Machine Learning**

Artificial intelligence (AI) is concerned with the study and design of systems that can perform complex actions, such as translating a text, driving a car, blocking unwanted emails, or recommending a book. An important type of action routinely performed by intelligent software is that of classification: assigning an item to one of many possible categories. This includes, as a special case, the task of making discrete decisions:[11] for example, about blocking or recommending a web page, or diagnosing a patient. While decisions can be made based on various mechanisms, modern AI methods are often created by exposing a learning algorithm to large numbers of training examples, a procedure known as 'machine learning'.

In the language of AI, the properties used to describe the item to be classified are often called "features" and the classes that are assigned to each item are called "labels". So, we could describe an email by the words it contains, a patient by the outcome of a set of clinical measurements, a client by a set of parameters describing their track record of payments. The labels applied to an email can be, for example, 'spam' or 'ham', those applied to a patient could be 'diabetic' or 'healthy', those for the loan-client could be 'safe' or 'risky'. The goal of a learning algorithm is to build a function (a "classifier") that assigns a class label (e.g. 'spam' or 'ham') to any object (e.g. emails) that has not yet been labelled.

---

[11] The output of a classification algorithm might be considered a decision since it results from choosing among a set of categories or a score that puts an item within a certain category. This should be distinguished from the consequential decision that usually humans make after consulting algorithmic output (e.g. doctor's decision of requiring further screening to a patient based on the algorithmic output classifying that patient into a "malignant tumour" category).

**Performance of Classifiers**

The performance of any given classifier is defined in terms of the mistakes it makes on new and unseen data (called a "test set"), and it can be estimated on a dataset of known cases (called "training set"). For example, given a set of emails already labelled as 'spam' or 'ham', how many errors does the classifier make, of each type, when it is run on a set of unlabelled emails? In this two-class example, mistakes can be either false positives (e.g. 'ham' emails mistakenly classified as 'spam') or false negatives (e.g. 'spam' emails incorrectly labelled as 'ham') and can have different costs (deleting a valid email might be more costly than keeping an unwanted one). The confusion matrix (see table 1) is a data structure that keeps track of the four possible outcomes:

|  | **True spam** | **True ham** |
|---|---|---|
| **Predicted spam** | True Positive (TP) | False Positive (FP) |
| **Predicted ham** | False Negative (FN) | True Negative (TN) |

*Table 1: 2x2 confusion matrix*

Future performance is what matters when deploying such a system, and this is where significant work has been done in the theory of machine learning: what can past performance tell us about future performance? This is typically quantified in terms of probability: the probability of a given type of mistake being made, under certain (controllable) conditions.

In home pregnancy tests the labels are 'pregnant' and 'non-pregnant' and only one feature is used: the concentration of certain hormones in urine. The test can lead to false positives and false negatives, each with a different cost (Is it better to miss a pregnancy or to raise a false alarm? This depends of course on the usage situation). Past performance of this test on a sample of patients can give us information about its future performance, but only under the strong assumption that the subjects were selected in the same way. For example, if we test it on teenagers and then we use it on middle-age subjects, it might not have similar levels of performance.

To quantify the probability that a certain type of mistake is made in the future, we need to make some assumptions about the source presenting the future items to the classifier, as mistakes can occur with different probability. A typical statement that can be made in statistical learning is: if the future data are sampled according to a distribution of probability which is the same as that used in the test phase, then with high probability the error-rates will be close to those observed in the test phase.

A technical detail that will be relevant later in this article is that many (but not all) binary classifiers work in two steps: first a real valued score is computed for the item to be classified, then that score is compared with a threshold, or cut-off point, and the item is assigned to one class or the other, based on where it falls. That real valued score could informally be thought of as a probability, though it is not necessarily formally a probability. For this specific case, there is an extra consideration: by moving the threshold we have a trade-off between false-positives and false-negatives. In the clinical example of pregnancy testing this could be captured by the notions of sensitivity or recall (i.e. the true positive rate: TP/(TP+FN)) and specificity (i.e. the true negative rate: TN/(TN+FP).

Other common performance measures are: Overall accuracy ((TP+TN)/(TP+FP+TN+FN)), which tell us how often the algorithm classifies items correctly, and precision (i.e. the positive predicted value: TP/(TP+FP)), which estimates the fraction of relevant positive instances returned by the algorithm. For our case study other relevant metrics are the probability of mistakes: the false discovery rate (i.e. the expected proportion of discoveries which are false, FP/(TP+FP)) and the false omission rate (i.e., the expected proportion of discoveries which are omitted: FN/(FN+TN)).

While this threshold must be set before the system is deployed, when engineers compare the merits of different scoring methods they often compare all possible error-rates for all possible thresholds: in other words, by moving the threshold from the minimum to the maximum level, they generate pairs of true-positives rate (sensitivity) - false positive rate (the probability of false alarm) that are plotted on a diagram called a receiver operating characteristic (ROC) curve, whose mathematical properties are well known.

ROC analysis can add further insights into model performance enabling the algorithm designer to visualize different trade-offs between the benefits (the true positive rate) and the costs (the false positive rate) of a classifier. Often ROC performance is reduced to a scalar value, the so-called area under the curve (AUC), which can range between 0 and 1, where 1 indicates a perfect classification and 0.5 a random guess.

**Learning to Classify**

In most practical cases, the designers of an AI system decide the features that are used to describe an item, and the classes that are available, but they do not design the actual decision function. This is automatically generated by a learning algorithm which has access to many training examples: large sets of labelled-data which can guide the learning algorithm towards the best configuration of parameters, to maximise some measure of performance. This is why the design of automated classifiers is really a branch of machine learning,[12] and also why training data is such a valuable commodity.

It is important to notice that the system designer makes another important choice: the class of possible decision functions that can be output by the system. This may be a very large space, which the learning algorithm will have to explore, guided by the training data, but will necessarily be a subset of all possible decision functions.

As the decision function has been selected by an algorithm based on training data, the assessment of its performance on new unseen data is particularly important. Statistical Learning Theory[13] provides guidance for this step: if a classifier is tested on data that was never seen during its training phase and is of sufficient size, and it is found to perform well, then we can expect it to perform well on future data that was obtained in the same way as the test set.

The crucial points above are: a) the test set needs to be of sufficient size; b) it should not have been used for training; and c) the future data must be obtained in the same way as the test set. Violating those requests means that we do not have a reliable assessment of the future performance of the classifier.

For the spam filter, a certain combination of words might appear to be associated with spam emails, but on a small test sample this might also be a coincidence. This kind of coincidence becomes more probable if the test data was already seen in the training phase (a problem known as 'overfitting'[14]), or - equally - if changes are made to the classifier in response to the test data. Finally, even if the classifier performs well on a large number of new and unseen test emails, its performance cannot be guaranteed if it is applied to a different source of data. For example, if something is found to be effective in detecting spam in emails from 1995, it might still fail on emails from 2017.

It is worth noting that there are many types of functions that can be used to map features to classes. The most common ones are based on linear functions (class is predicted based on a score that is a linear combination of features), neural-networks (the class-score is a nonlinear function of features), decision-trees, and various types of combinations and committees of those. Interpreting how the classification function combines the various input features to compute a decision is not an easy task, in general, and goes beyond the boundaries of Statistical Learning Theory.

---

[12] This is the automation of statistical work previously done by actuarial-scientists, statisticians, etc. There is a lot of overlap - but with machine learning this happens on a very large scale and in a largely autonomous manner, and the effects of the scale, automation, and future development of the field is the concern of this article.

[13] Statistical Learning Theory is a mathematical framework for the field of machine learning which became popular with the development of the so-called Support Vector Machines (Vapnik 1995)

[14] Overfitting is the result of a statistical model which performs well over the training sample but generalizes poorly to new data. Underfitting occurs when the model does not capture the underlying structure of data (e.g. when a linear model is applied to non-linearly separable data).

**Correlation vs. Causation**

These important effects (like overfitting) are all related to the key fact that, at present, classifiers make predictions based on statistical correlations discovered between the features and the labels associated with the training items, not on any causal relations. So, it is possible that a statistical relation may disappear if the sampling distribution is changed, or that a spurious relation may appear as the result of what statisticians call "multiple testing" – i.e. when testing multiple hypotheses simultaneously which are actually non-significant but can produce a significant result just by chance. The scoring function will not capture the essence of what makes a spam email but will be able to predict - and bet - based on indirect clues that associate with the labels.

To summarise, overfitting and out-of-domain application are both the result of the system relying on associations that appeared to be informative in the training or evaluation stage but are no longer informative during the deployment of the system.[15]

However, this is also part of the success of machine learning: predictions can be made based on combinations of features that somehow end up correlating with the target signal. This is the way in which purchase behaviour might be informative about voting intentions,[16] and therefore valuable.

When the machine cannot explain the reasons behind a guess, we cannot interpret its motivations, we can only assess its input-output, we talk about a 'black box' - the email might have been removed, but we will not know why. In other words, machine learning predictions or classifications are educated guesses or bets, based on large amounts of data, and can be expected to work subject to certain assumptions.

Often the specific blend of features that happens to correlate with the target signal has no real meaning to us. For example, a home pregnancy test is not based directly on the cause of pregnancy, but on correlating hormonal factors. However, what if we had a pregnancy test based on email content? The frequency of certain expressions might suggest an increased probability of pregnancy (e.g. morning sickness) in the general population, but this would not be causal, and the application of this general rule to a specific population - say medical students' inboxes - might fail entirely.[17]

**On Bias**

The fact that we trust certain correlations, or a class of possible decision functions, largely depends on a series of assumptions and inductive bias (Mitchell 1997). The term 'bias' is used with different meaning in different contexts. Informally, 'bias' is used to indicate any deviation from neutrality (or uniformity), any preference (or preconception), but its technical use in probability, statistics, machine learning and social sciences can be very specific. 'Bias' is not used in a pejorative way in the STEM literature.

For example, in the theory of probability the term 'bias' is often used to indicate the probability associated with a binary event such as a coin toss. The bias of a coin is its probability of landing on head, and a coin is called 'unbiased' or 'fair' when that probability is 1/2.

---

[15] Overfitting was one of the reasons that led Google Flu trends to predict "more than double the proportion of doctor visits for influenza like-illness (ILI) than the Centers for Disease Control and Prevention (CDC)" (Lazer *et al.* 2014: 1203)

[16] E.g. according to Kantar UK Insights: https://uk.kantar.com/ge2017/2017/what-your-shopping-basket-says-about-how-you-vote/: " Conservatives buy more products from the alcohol and fresh fish categories" whereas "Labour supporters put more toiletries into their trolley".

[17] However, consider the notorious case where Target, the US discount store, was able to predict a teenage girl was pregnant, before her own family knew, based on the analysis of historical purchase data for women who had signed up for Target baby registries in the past (Duhigg 2012). Purchase data were correlated to pregnancy and so they were able to predict it but, of course, the data played no causal role.

In statistics, where we usually estimate a property of a distribution, the notion of bias refers to an estimator.[18] If the expected value of that estimator is the same as the true value of the quantity being estimated, the estimator is said to be unbiased. Sometimes biased estimators are preferred to unbiased ones. This occurs when biased estimators, for increasing sample size, converge to the true value faster than unbiased ones.

In machine learning a related concept is used: learning a concept from a finite sample requires making some assumptions about the unknown concept, so as to reduce the search space, and reduce the risk of overfitting the training set. This is done by knowingly introducing a bias in the system, that is a preference for a certain type of outcome.[19] For example, we can force a spam filter to only use linear combinations of word frequencies, or to prefer simple decision rules, thus reducing the options of the learning algorithm. Occam's razor, the principle that the simplest hypothesis is to be preferred, all else being equal, is a classic example of bias in machine learning.

**MACHINE DECISIONS IN CRIMINAL JUSTICE**

**Risk Assessment Tools**

The development of risk assessment tools has an established tradition in criminal justice and, since the early eighties, actuarial methods have been used to inform decision-making in correctional institutions (Feeley & Simon 1994). In the justice system, risk assessment tools are used to measure the future behavioural risk posed by a defendant and to inform a variety of decisions at different steps of the criminal justice process and in several types of crime. For example, risk assessment tools have been deployed in pretrial release,[20] in parole and probation[21] (i.e. to identify the most appropriate supervision) and criminal sentencing.[22] There are also tools developed for specific types of crimes, such sex offences,[23] youth delinquency[24] or domestic violence.[25]

In this section we will review the Harm Assessment Risk Tool (Urwin 2016 and Barnes 2016), as a recent example drawn from a huge literature[26] and, most importantly, as a concrete application of machine decision-making. In the light of the technical descriptions outlined in the previous section, we will try to point out the main components of the tool, i.e. the classification method, the features used, the training and the test sets, the implied assumptions and the performance measures.

**The Harm Assessment Risk Tool (HART)**

---

[18] E.g. if we are estimating the expectation of a random variable based on a finite sample, the estimator might be the average of that sample.

[19] It is known in machine learning that without such a learning bias the general problem of learning from finite data is ill-posed. See the so-called "inductive learning hypothesis" (Mitchell 1997).

[20] The Public Safety Assessment tool, developed by Laura and John Arnold Foundation (http://www.arnoldfoundation.org/) is used by 21 jurisdictions including three entire states, i.e. Arizona, Kentucky and New Jersey (Christin *et al.* 2015)

[21] The Correctional Offender Management Profiling for Alternative Sanctions (COMPAS) which has been developed by a for-profit company, Northpointe (now Equivant), is one of the most popular risk assessment tools in USA (http://www.equivant.com/solutions/inmate-classification)

[22] Pennsylvania is in the process of extending the use of risk assessment tools for sentencing decisions (Barry-Jester *et al.* 2015) and, interestingly, also the state of Wisconsin is using COMPAS for sentencing (Tashea 2017).

[23] Static-99/R is the most widely used sex offender risk assessment instrument in the world and is used extensively in the United States, Canada, the United Kingdom, Australia, and many European nations. For more details see the website: www.static99.org

[24] Many risk assessment tools offer versions for young offenders, such as COMPAS (http://www.equivant.com/solutions/case-management-for-supervision) and the Violence Risk Scale (http://www.psynergy.ca/VRS_VRS-SO.html).

[25] The Ontario Domestic Assault Risk Assessment (ODARA) predicts the risk of future domestic violence http://odara.waypointcentre.ca/ and is available for police, victim support and health services. It was developed by the Ontario Provincial Police and the Research Department at Waypoint. Other similar tools are the Violence Risk Appraisal Guide Revised (see the official website http://www.vrag-r.org/) and the Domestic Screening Violence Inventory (see Williams & Grant 2006) .

[26] See examples in footnotes 4 and 15-20.

Launched in May 2017 to extensive media coverage(e.g. Baraniuk 2017, Sulleyman 2017), the Harm Assessment Risk Tool (HART) is an application developed by Durham Constabulary and Cambridge University to support police officers with custody decisions[27]. The model predicts the likelihood that an offender will commit a serious offence (high risk), non-serious offence (moderate risk), or no offence (low risk), over a two-year period after the current arrest.

**The Classification Method**

HART is built using random forests, a machine learning method that results from the combination of hundreds of decision trees (Oswald *et al.* 2018 and Urwin, 2016). A decision tree is a popular algorithm where classification can be thought of as a series of if-then rules (Mitchell 1997: 52). Each node of the tree tests some attribute of the item to be classified (e.g., age, gender, years in jail, etc.) and each descending branch represents a possible value for the considered attribute/node (e.g. for gender there might be two values: "female" or "male", for age there might be some predefined intervals such as "0-18", "19-30", "31-50" and so on). Starting from the root, each instance is sorted at each node based on the value taken by the specified attribute moving down until a leaf, i.e. the terminal node containing a class label, is reached. Usually, we refer to classification trees when the labels take discrete values ('Yes' / 'No' answers or integers) while we talk about regression trees when the labels take continuous values (usually real numbers).

Random forests construct a multitude of decision trees trained on random subsamples of the training set and using a random subset of features. Predictions of each individual tree can be averaged (with regression trees) or aggregated by taking the majority vote (in the case of classification trees). HART consists of 509 decision trees, each one producing a separate forecast that counts as one vote out of 509 total votes and outputs the forecast that receives the most votes (Oswald *et al.* 2018 and Urwin, 2016).

A key feature of random forests is the out-of-bag error, which provides an estimate of the generalization error during the training phase. When a random sample is drawn to grow a decision tree a small subset of data is held out (usually one third of the sample) and used as a test set of that tree. This process, commonly known as "Out of Bag (OOB) sampling", is used internally, during the run of the model, to estimate an approximation of model accuracy.

Another property is the ability to balance different types of errors, so that mistakes with highest costs occur less frequently than those less costly. For example, classifying a future serious offender into a lower risk category (false negative, i.e. someone classified as low-risk who commits a crime in the follow-up period) is considered costlier than misclassifying a non-serious offender into a high-risk category (false positive, i.e. someone flagged as high risk who does not commit any crime in the follow-up period)[28]. In HART these two errors are termed "dangerous error" and "cautious error" respectively (Oswald *et al.* 2018: 227 and Urwin 2016: 43) and their costs vary accordingly.

As Oswald *et al.* (2018) pointed out, the set-up of cost ratios is a policy choice that was taken prior to the model's construction with the involvement of senior members of the Constabulary. In these respects, HART intentionally prefers (i.e., applies a lower cost ratio to) false positives (the so-called cautious error) over false negatives (the so called dangerous errors). In this way, "the model produces roughly two cautious errors for each dangerous error." (Oswald *et al.* 2018: 227)

**The Training and the Test Sets**

The training dataset is composed of 104,000 custody events from a period between January 2008 and December 2012. A custody event is defined as "the disposal decision taken by the custody officer following arrest at the end of the first custody period" (Urwin 2016: 37). The output of a custody decision could be: "to bail (conditional or unconditional), remand in custody, taken no further action, administer an out of court disposal/diversion scheme or prosecute the suspect with a decision to bail (conditional or unconditional)" (Urwin 2016: 37)[29].

---

[27] HART was introduced as part of the Checkpoint programme, an initiative within Durham Constabulary which aims to reduce reoffending by offering moderate risk offenders an alternative to prosecution. See (Oswald *et al.* 2018) and Durham Constabulary website: https://www.durham.police.uk/Information-and-advice/Pages/Checkpoint.aspx)

[28] From a technical point of view a way to build the costs of classification errors is to alter the prior distribution of the outcomes so as to reflect the desired policy. For example, suppose that 30% of the offenders in the training set "fail" and 70% do not. By this procedure the prior proportions can be changed so that failures are made relatively more numerous (Berk 2012).

The model was validated in two steps. A preliminary validation has been made with the aforementioned "Out of Bag sampling" procedure, thus with random samples of the construction (training) dataset. A more recent validation (Urwin 2016) has been conducted by using a separate test set composed of 14,882 custody events occurred in 2013. All data sets were drawn from Durham Constabulary management IT system.

**The Features and the Labels**

The training set includes 34 features regarding offenders' demographics and criminal history (see table 2). Note that the feature set also includes two types of residential postcodes. The former ("CustodyPostcodeOutwardTop24") contains the first four characters of the offender's postcode, while the latter ("CustodyMosaicCodeTop28") can take a value from 28 available codes providing information about most common socio-geo demographic characteristics for County Durham.[30]

While acknowledging the ethical implications of including offender's postcode among predictors – this could give rise to a feedback loop that may reinforce possible existing biases, so that people living within areas targeted as "high risks" would undergo closer scrutiny than those living in other areas (more will be said on the dimension of fairness in the following section) - Oswald *et al.* (2018) argue that the nature of random forests would prevent a single feature, like postcode, from determining the entire forecasted outcome[31]. Yet, Oswald *et al.* (2018) reported that one of the two postcodes could be removed in a later iteration.

The model uses 3 labels reflecting the classification of offenders: 1) high risk = a new serious offence within the next 2 years; 2) moderate risk = a non-serious offence within the next 2 years; 3) low risk = no offence within the next 2 years.

| FEATURE | DESCRIPTION |
|---|---|
| 1. CustodyAge | Age at presenting custody event |
| 2. Gender | Male or female |
| 3. InstantAnyOffenceCount | Count of any offences at presenting custody event |
| 4. InstantViolenceOffenceBinary | A yes/no binary value is used to define the present offence in terms of violence |
| 5. InstantPropertyOffenceBinary | A yes/no binary value is used to define the present offence in terms of property offence. |
| 6. CustodyPostcodeOutwardTop24 | The 25 most common 'outward' (first 3-4 characters) postcodes in County Durham. If the offender's |

---

[29] Note that the custody events of the training and the test sets are supposed to regard decisions with an observable outcome (e.g. "released on bail") – indeed for those suspects who remain in jail we have no observation.

[30] According to Big Brother Watch (2018) the second postcode feature refers to Experian's Mosaic data. Mosaic UK is Experian's tool providing a classification of UK households relating sociodemographic characteristics to postcodes (http://www.experian.co.uk/marketing-services/products/mosaic-uk.html). The Mosaic system generates information about typical individuals and families in postcode areas (common social types are: "World-Class Wealth", "Disconnected Youth", "Low Income Workers", etc.) Widely used for marketing purposes, it can also be used for credit scoring and crime mapping.

[31] Indeed, "algorithms such as random forests are based upon millions of nested and conditionally-dependent decisions points, spread across many hundreds of unique trees [...] It is therefore the combination of variables, and not the variable in isolation, that produced the outputted risk level" (Oswald *et al.* 2018: 228).

|  |  | postcode is outside of County |
| --- | --- | --- |
| 7. | CustodyMosaicCodeTop28 | The 28 most common socio-geo demographic characteristics for County Durham |
| 8. | FirstAnyOffenceAge | The suspect's age at first offender regardless of juvenile or adult |
| 9. | FirstViolenceOffenceAge | The suspect's age at first violent offence regardless of juvenile or adult |
| 10. | FirstSexualOffenceAge | The suspect's age at first sexual offence regardless of juvenile or adult |
| 11. | FirstWeaponOffenceAge | The suspect's age at first weapon offence regardless of juvenile or adult |
| 12. | FirstDrugOffenceAge | The suspect's age at first drug offence regardless of juvenile or adult |
| 13. | FirstPropertyOffenceAge | The suspect's age at first property offence regardless of juvenile or adult |
| 14. | PriorAnyOffenceCount | The number of offences prior to the presenting offence for the suspect |
| 15. | PriorAnyOffenceLatestYears | The number of years since any offence– if there is no offence history, Null value is returned |
| 16. | PriorMurderOffenceCount | The number of murder offences prior to the presenting offence for the suspect |
| 17. | PriorSeriousOffenceCount | The number of serious offences prior to the presenting offence for the suspect |
| 18. | PriorSeriousOffenceLatestYears | The number of years since the most recent custody instance in which a serious offence was committed – if there is no serious offence history then a code of 100 years is used. |
| 19. | PriorViolenceOffenceCount | The number of violence offences prior to the presenting offence for the suspect |
| 20. | PriorViolenceOffenceLatestYears | The number of years since the most recent custody instance in which a violence offence was committed – if there is no violence offence history then a code of 100 years is used. |
| 21. | PriorSexualOffenceCount | The number of sexual offences prior to the presenting offence for the suspect |
| 22. | PrioriSexualOffencelatestYears | The number of years since the most recent custody instance in which a sexual offence was committed – if there is no sexual offence history then a code of 100 years is used. |
| 23. | PrioriSexRegOffenceCount | The number of sex offender register offences prior to the presenting offence for the suspect |

| 24. PrioriWeaponOffenceCount | The number of weapon offences prior to the presenting offence for the suspect |
|---|---|
| 25. PriorWeaponOffenceLatestYears | The number of years since the most recent custody instance in which a weapon offence was committed – if there is no weapon offence history then a code of 100 years is used |
| 26. PriorFirearmOffenceCount | The number of firearms offences prior to the presenting offence for the suspect |
| 27. PrioriDurgOffenceCount | The number of drug offences prior to the presenting offence for the offender |
| 28. PriorDrugOffenceLatestYears | The number of years since the most recent custody instance in which a drugs offence was committed – if there is no drugs offence history then a code of 100 years is used |
| 29. PriorDrugDistOffenceCount | The number of drug distribution offences prior to the presenting offence for the offender |
| 30. PriorPropertyOffenceCount | The number of property offences prior to the presenting offence for the offender |
| 31. PriorPropertyOffenceLatestYears | The number of years since the most recent custody instance in which a property offence was committed – if there is no property offence history then a code of 100 years is used. |
| 32. PriorCustodyCount | The number of custody events prior to the presenting offence for the offender |
| 33. PriorCustodyLatestYears | The number of years since the most recent custody instance – if there is no custody event history then a code of 100 years is used |
| 34. PriorIntelCount | The number of intelligence submissions at nominal level. The offender at nominal level will have a unique identifier, the submissions are counted within the forecasting model |

*Table 2: List of features used to train HART (Urwin, 2016)*

**Validation and performance measure**

In Urwin's validation study (2016), HART's performance has been assessed with respect to both the Out of Bag samples (OOB construction data) and the separate test set (2013 validation data). For each dataset a 3 x 3 confusion matrix (see table 3 and 4) keeps track of the standard error measures expressed in percent values and conditional to the 3 possible labels (high-risk / moderate-risk / low-risk).[32]

| OOB Construction data | Actual High | Actual Moderate | Actual Low | Total |
|---|---|---|---|---|
| **Forecast High** | 8.12% | 6.80% | 1.80% * | 16.72% |
| **Forecast Moderate** | 2.25% | 34.09% | 12.19% | 48.53% |
| **Forecast Low** | 0.82% + | 7.65% | 26.28% | 34.75% |
| **Total** | 11.18% | 48.54% | 40.27% | 100% = 104,000 custody events |

*Table 3: Confusion Matrix for the training set (see table 7 in Urwin, 2016: 54)*

| 2013 Validation | Actual High | Actual Moderate | Actual Low | Total |
|---|---|---|---|---|
| **Forecast High** | 6.26% | 10.01% | 2.23% * | 18.49% |
| **Forecast Moderate** | 4.88% | 32.53% | 13.55% | 50.95% |
| **Forecast Low** | 0.73% + | 5.81% | 24.02% | 30.55% |
| **Total** | 11.86% | 48.35% | 39.79% | 100% = 14,882 custody events |

*Table 4: Confusion Matrix for the test set (see table 8 in Urwin, 2016: 54)*

Among cautious and dangerous errors two notions are distinguished (Urwin, 2016: 54-55):

- "very dangerous error" = a suspect being labelled as 'low risk' commits a serious offence in the next 2 years (see cells marked with "+" in table 3 and 4)
- "very cautious error" = a suspect being labelled as 'high risk' does not commit any crime in the next 2 years. (see cells marked with "*" in table 3 and 4)

In the table below, we report some performance measures of HART (see table 5 which was created by combining tables 6 and 9 in Urwin, 2016: 52, 56) – note that they can be easily computed by using the terminology introduced in the previous section of this chapter.

---

[32] Note that the validation study also assessed to what extent the custody officers' decisions agree with HART predictions. Using a separate data set consisting of 888 custody events for the period: 20 Sep. 2016 - 9 Nov. 2016), the study suggested that the highest level of agreement is in the moderate category, at 39.86%, while clear disagreement emerges in the high-risk category – the police and the model agree only at 1.58% (Urwin 2016: 71-72).

|  | OOB construction data | 2013 validation data | |
|---|---|---|---|
| **Overall accuracy**: what is the estimated probability of a correct classification? | 68.50% | 62.80% | |
| **Sensitivity / recall**: what is the true positive rate for each class label? | 72.60% | 52.75% | HIGH |
| | 70.20% | 67.28% | MODERATE |
| | 65.30% | 60.35% | LOW |
| **Precision**: what is the rate of relevant instance for each class label? | 48.50% | 33.83% | HIGH |
| | 70.20% | 63.84% | MODERATE |
| | 75.60% | 78.60% | LOW |
| **Very dangerous errors**: of those predicted low risk, the percent that was actually high risk (subset of the false omission rate) | 2.40% | 2.38% | |
| **Very cautious errors:** of those predicted high risk, the percent that was actually low risk (subset of the false discovery rate) | 10.80% | 12.06% | |

*Table 5: some performance measures of HART extracted from tables 6 and 9 in Urwin (2016: 52,56)*

To the best of our knowledge, ROC analysis and the AUC value are not provided in the validation study (Urwin 2016).

**OPERATIONALISING TRUST: FOUR NORMATIVE BENCHMARKS FOR MACHINE DECISION-MAKING**

As noted above, HART is a case study drawn from a wider reality and, in addition to criminal justice, many other domains are reframing routine decisions in machine learning terms (education, recruitment, health care, government, etc). This reframing is premised on the fact that machine learning captures the general mechanism underlying common "diagnostic questions" (Swets *et al*, 2000): Will this person commit violence? Will this student succeed? Will this candidate meet current and future company's needs? Which costumers will best match your business? Which persons will be more likely to develop depression? Which citizens should be prioritised in tax exemption? Usually all these questions call for a positive or negative decision about the occurrence of future event (student's success, employee's achievements, the rise of depression…) or the presence of specific conditions (e.g. for accessing a tax reduction programme or other public services).

Many argue that a thoughtful application of machine learning to (diagnostic) decisions might improve public policy (Kleinberg *et al.* 2016). But, while the study of the effective advantages of machine decisions need further investigation,[33] the overall assessment of these technologies should extend beyond accuracy guarantees. When machine learning software evaluates an individual for the purposes of making a decision about their entitlement to some tangible benefit or burden, the operating procedures and consequences of their use should be evaluated according to the principles and values of democratic society.

---

[33] E.g. on one hand, Kleinberg *et al.* (2017) showed that machine learning could provide important economic and welfare gains in bail decisions, on the other, Dressel & Farid (2018) revealed that machine learning algorithms (e.g. COMPAS), when applied to pretrial decision-making, are no more accurate and fair than untrained (Mechanical Turk) workers. Urwin (2016) also addressed the problem of human vs. machine predictions, but the results of such analysis are not yet available as the forecasts were made in 2016 and the follow-up period has not yet passed (see the written evidence submitted by Sheena Urwin, Head of Criminal Justice at Durham Constabulary, in relation to the enquiry into Algorithms in Decision-Making of the Science and Technology Committee, House of Commons, UK: http://data.parliament.uk/writtenevidence/committeeevidence.svc/evidencedocument/science-and-technology-committee/algorithms-in-decisionmaking/written/78290.pdf) .

In the following sections we will examine four benchmarks derived from the normative principles/objectives discussed above that may help stakeholders, including practitioners and citizens, answer the following question: How does one create an effective trust framework (e.g. one that permits rational development of trust and efficient exercise of distrust) that will legitimise the augmentation, or replacement, of human decision-making by machine decision-making? These benchmarks are derived from established normative principles and reflect the primary factors in the contemporary debate surrounding the quality of machine decisions. We believe these could form the basis of minimum requirements for legitimating the use of machine decision-making and, hence, set a standard for current and future applications. The benchmarks are explored here in the light of the HART case study, but their use as an analytical tool to inform regulatory theorising is extensible beyond the criminal justice sector, and has potential for cross-domain application.

**Prediction Accuracy**

As discussed above, there are many ways to quantify the performance of a classifier and various aspects to consider when model accuracy is reported. Two important pieces of information are: which performance measure is used (e.g. overall accuracy? 'area under the curve' (AUC)?); and how was this measured (how large was the test or validation set; how was it formed?). Let us consider these questions in the light of our running example.

HART was validated by using two datasets: the construction sample (used also for training) and the validation sample (not used for training). The validation study provided several measures such as overall accuracy, sensitivity, precision and the expected probability of two specific mistakes, i.e. "very dangerous" and "very cautious errors".[34]

When comparing model performances with the two datasets (i.e. training and test sets) the classification accuracy declines (see Urwin 2016:50-52). This holds for both overall accuracy which falls from 68.50 % (in the OOB construction data) to 62.80% (in the 2013 validation data), and sensitivity (the true positive rate) which gets worst with the high-risk category (it passes from 72.60% to 52.75%). So, one might ask whether using HART provides real advantages (greater accuracy) with respect to a more trivial decision mechanism (the so-called "baseline"): Would it perform better than a random guesser? Computing the prediction of a random baseline we observe that HART sensitivity would be definitely more accurate. If the percent of actual outcomes is known the accuracy of a random baseline can be calculated by using the rules of probability.

We denote Y as the actual outcome and Ŷ as the prediction of a random guesser which is picked from the population of the 2013 validation dataset (comprising 11.86 % of high risk, 48.35% of moderate risk and 39.79% low risk offenders). Then the accuracy of the random guesser is:

$$[P(Y = \text{"high"}) * P(Ŷ = \text{"high"})] + [P(Y = \text{"moderate"}) * P(Ŷ = \text{"moderate"})] + [P(Y = \text{"low"}) * P(Ŷ = \text{"low"})] =$$
$$[0.1186 * 0.1186] + [0.4835 * 0.4835] + [0.3979 * 0.3979] = 0.406 = 41\%$$

The decline of accuracy might relate to the differing of offender samples (i.e. OOB construction sample vs. 2013 validation sample) – indeed the frequency of serious crimes increased between 2013 and 2015.[35] Thus, it might be the case that HART captured statistical relations that were more effective in classifying offenders in the training set, but that were less significant in the test set.

As we have seen before, the statistical associations found in the training phase might also become less informative because of an out-of-domain application, i.e. when sampling conditions vary. In the case of criminal justice, we might ask: Are judges' decisions made in the same way? Did they predict the same outcome? have the regulatory bodies practices changed (rules, police routines, etc.)? For

---

<br>

| | |
|---|---|
| 34 | For an extensive account of the metrics computed, see Urwin (2016) |
| 35 | See the percentages of recorded crimes in Urwin (2016: 47). |

instance, a risk assessment tool trained on a population of parolees might not be appropriate to inform bail decisions (the two populations have different characteristics: usually parolees have already spent time in jail) and thus provide poor or wrong predictions (Berk 2012). Note that in the case of HART these problems might be kept under control if, as Oswald *et al.* (2018) reported, the data are drawn only from the Durham Constabulary system and not from other sources, such as local agencies in Durham or national IT systems.

The characterization of a model's performance, as we have noted above, depends also on the fact that not all mistakes are equal: for binary classifiers, one would need to separate the effect of false positives and false negatives; for multi-class classifiers, there would be many combinations and types of error, potentially each with different cost. For example, HART's developers have deliberately weighted more false negatives than false positives. This, as Oswald *et al.* (2018) stressed, is a policy choice rather than a design strategy reflecting specific social needs and preferences (e.g. that of safeguarding citizens from potential harms). This choice is reflected in HART's performance: The percent of very dangerous errors is lower than that of very cautious errors and the former remains the same in both datasets (very cautious errors slightly increase).[36]

**Fairness and Equality**

The findings of *ProPublica*'s analysis (Larson *et al.* 2016) suggested that the use of risk assessment tools might bring about social discrimination.[37] Social disparities are not new in on-line advertising (Sweeney 2013, Datta *et al.* 2015) and computer systems (Friedman & Nissenbaum 1996).

All these findings naturally raise questions for machine decision-making: can AI systems discriminate? How do they do it? Which technical, legal and social remedies exist to avoid algorithmic discrimination? Which criteria can help us to inform such remedies?

A basic intuition underlying the notion of fairness is that all human beings must be treated equally before the law. This idea is deeply rooted in human rights documents which acknowledge that "all are entitled to equal protection against any discrimination" (UN Universal Declaration of Human Rights, Article 7). Generally, we consider fairness as a comparative notion:[38] what counts for a person is to be treated equally with respect to other people who are subject to the same procedure. Thus, when a person or a group is unjustifiably differentiated and, because of this, put in unfavourable conditions, they are discriminated against.

What discrimination is and how it occurs is a controversial issue. Often the law, rather than providing a definition, prefers to indicate a non-exhaustive list of prohibited attributes that cannot be used to ground decisions in various settings (such as housing, education, employment, insurance, credit, etc.). Typically, the list includes: race, colour, sex, religion, language, political or other opinion, national or social origin, property, birth or other status.[39] These attributes are then used to characterise groups of people which are recognised as important to the structure of a society and, hence, worthy of a special protection ("protected groups").

---

[36] Specifically, "the model was able to ensure the likelihood of very dangerous errors occurring was just 2%. Therefore, the organisation can be 98% sure that a very dangerous error will not occur." (Urwin, 2016: 85). Technically speaking, this result says that 98% of the time that the model predicts somebody as low risk, they will not commit a serious crime (see the Cambridge Centre for Evidence-Based Policing video: https://www.youtube.com/watch?v=zc8x5P7suuo&feature=youtu.be) Hence, she/he will be a moderate or low-risk offender. This is not the same as the media claim that "forecasts that a suspect was low risk turned out to be accurate 98% of the time" (Baraniuk 2017).

[37] In short, the analysis found that COMPAS made more mistakes with black defendants as compared to their white counterparts, i.e. the false positive rate was higher with black defendants and the false negative rate was higher with white defendants.

[38] For an alternative view, see also Hellman (2016).

[39] These characteristics are referred to, in the context of non-discrimination, in international human rights documents such as the International Covenant on Civil and Political Rights (Article 26), and European Convention on Human Rights (Article 14). Most are also to be found in international data protection agreements, e.g. The Council of Europe's Convention for the Protection of Individuals with regard to Automatic Processing of Personal Data (Article 6), and the EU General Data Protection Regulation (Article 9), as personal data to which special safeguards should be applied.

Algorithmic discrimination[40] can arise at different stages of the design process (Barocas & Selbest 2016). For instance, when forms of social bias are incorporated in the training data, the classifier is likely to reproduce the same bias (see, for example, Caliskan *et al.* 2017 for the case of language).[41] Other pitfalls might be found during the feature selection process - when sensitive attributes in the feature set correlate to a classification outcome (e.g. when gender attribute is systematically associated to lower paid jobs)[42] - and sampling - when certain protected groups are underrepresented as compared to others.[43]

Building fair (or non-discriminating) algorithms supposes the implementation of an equality principle which can be intended in various ways. From a purely formal point of view, equality is achieved when alike cases are treated as alike[44]. From a social perspective, equality is concerned with the distribution of goods and services. Its meaning has been specified in two competing conceptions:

- *Equality of outcomes*: everyone should receive the same levels of good or services. A straightforward example is the ideal that all receive the same income.

- *Equality of opportunities*: everyone should have the same chances of success irrespective of sensitive attributes or those that are irrelevant for the task at hand.

These principles inspire different fair practices. For example, a system of quotas (e.g. jobs for specific groups) would enforce equality of outcomes; a system that screens CVs only based on relevant attributes would enforce equality of opportunities. The same holds true also for the development of non-discriminating algorithms where different fairness criteria relate to different notions of equality.

A technical measure that relates to equality of outcomes is *statistical or demographic parity*. The latter requires that the portion of individuals classified as positive (e.g. "hired" or "high risk offenders") is the same across the groups (e.g. 'male' and 'female'). However, statistical parity is subject to several objections since ensuring the same portion of positive outcomes among groups does not guarantee the same level of accuracy, i.e. when there is an imbalance in the training set, the algorithm will learn and perform better with member of the overrepresented group (Dwork *et al.* 2012 and Hardt 2016). Moreover, a system of quotas might not be desirable in certain contexts such as that of criminal justice (Chouldechova 2017).

On the other hand, the implementation of equality of opportunity requires an in-depth consideration of how the classifier behaves across the groups (e.g. is the error rate the same across the groups? Is precision or sensitivity the same?). For example, popular metrics are:[45]

- *Calibration* which requires that the fraction of relevant instances among positive predicted outcomes (e.g. the fraction of predicted "high-risk" offenders which are indeed "high-risk") is the same across the groups. In other terms this notion requires equal precision across the groups.

---

[40] For a survey on this topic, see also Romei & Ruggeri (2013) and Zliobate (2015).

[41] Caliskan *et al.* (2017) have documented how standard natural language processing tools, which are trained on large corpora derived from the Web, reflect the same implicit bias found in language by the Implicit Association Test. For instance, European American names are more likely than African American names to be associated with pleasant terms - e.g. health, peace, honest, etc.- than unpleasant terms - e.g. abuse, murder, accident, etc.

[42] Note that a sensitive attribute can be redundantly encoded in other features. Thus, the algorithm may discover these encodings even if the sensitive attribute has been removed. For example, even though the gender attribute is dropped from the feature set, other attributes encoding gender (e.g. "cooking", "shopping", "family", "children") may correlate to jobs with lower salaries.

[43] If we have sample size disparities (i.e. a social group is overrepresented), the algorithm will learn statistical correlations that will badly transfer to the minority group (Hardt 2014). For example, if a credit score algorithm is trained on a dataset which disproportionally represents a segment of the population (e.g. "Caucasian individual with long credit history"), it could be error-prone on underrepresented groups (e.g. "Hispanic individuals with short credit history"). Similar effects were found in computer vision algorithms (see Dwoskin 2015 for a review).

[44] This principle has been formulated by Aristotle in the *Nicomachean Ethics*. For an extensive account of the notion of equality, see Gosepath (2011).

[45] For an extensive account of available metrics based on accuracy measures, see Berk *et al.* (2017).

- *Equalized odds* (Hardt *et al.* 2016) which, roughly speaking, enforces that the true positive rate is equally high in all groups. This corresponds to require equal levels of sensitivity and specificity in all groups.[46]
- *Error rate balance* (Chouldechova 2017) which requires that the classifier produces equal false positive and false negative rates across the groups. This is comparable to requiring equal misclassification rates (see the notion of disparate mistreatment in Zafar *et al.* (2016) and balance for positive and negative class (Kleinberg *et al.* 2017). Note that controlling the error rate balance is complementary to the notion of equalized odds.

Note that, all these notions, being comparative, require information about the size of the protected groups within the dataset (e.g. how many "women" and "men"). In the case of HART this information was not provided, and so no conclusion can be drawn.

Overall, the literature discussing these notions pointed out that it is impossible to satisfy multiple criteria at the same time. For example, Kleinberg *et al.* (2016) showed that there is no method to satisfy simultaneously calibration and the balance for positive class and the balance for the negative class. Similar conclusions were achieved by Chouldechova (2017), which considered the specific case of risk assessment tools, and Berk *et al.* (2017).

**Transparency and Accountability**

Transparency is a desirable property of legal and administrative systems. It can be understood as the right of citizens to access information about the procedures and data which lead to certain decisions affecting them. It is beneficial for several reasons: on the one hand, it can help organisations to account for the performed tasks, i.e. keeping track of the entire process that has led to a specific output (accountability), and, on the other, it can support citizens in understanding and questioning processes and outcomes (participation) .

While a right to transparency may not be explicitly expressed in human rights documents such as the UN Declaration of Human Rights and European Convention on Human Rights, it is arguable that without transparency as a penumbral, or background right, it would be difficult for an individual to ensure that those rights that are explicitly protected have been respected. Modern data protection legislation is explicit about the need to observe a principle of transparency in personal data processing (e.g. EU GDPR Recital 58 and Article 12; Council of Europe Draft Protocol amending the Convention for the Protection of Individuals with regard to Automatic Processing of Personal Data, Article 7).

The idea of transparency often implies an oversight body (e.g. audit commission or regulatory office) that can shed light on potential abuses or discriminations and, if necessary, motivate corresponding sanctions. For example, under the EU GDPR, ensuring compliance with transparency obligations will be part of the responsibility of an organisation's data protection officer, and will be subject to scrutiny by national supervisory authorities.

Usually transparency supports the design of accountable systems which allow one to detect responsibilities when some failures occur (e.g. when some records are lost or when they contain errors). Although transparency and accountability refer to distinct problems (what is called 'inscrutable evidence' and 'traceability' respectively in Mittelstadt *et al.* (2016) they are both directed at the diagnosis of potential or actual harms caused by a decision system. Indeed, when information is accessible and open to scrutiny it should be easier to understand where and why harms arise (i.e. doing system debugging) and, most importantly, who should be held responsible for them (Mittelstadt *et al.* 2016)

In the context of machine decisions, disclosing information may introduce new problems (Kroll *et al.* 2017). For example, if key factors, such as features and labels, are revealed, it may become possible to 'game the system' by seeking to alter variables that might influence the final output (see, e.g. search engine optimisation). Other obvious concerns arise from the publication of the dataset where sensitive information is used as input. However, the disclosure of the system's source code, while it may impact legitimate commercial interests (i.e. protecting a competitive advantage), does not necessarily imply transparency in the sense of intelligibility of the process or outcomes. Source code is usually incomprehensible to non-experts and even experts may not be able to predict algorithm behaviour just by examining the code (Kroll *et al.* 2017).

---
[46] In a binary classifier when the aim is to identify a "favourable" outcome (e.g. "admission" or "release", etc.) this notion is called "equal opportunity".

In addition, when approaching the transparency of machine decisions, we face another challenge that does not simply relate to the ability of reading a piece of code, but it refers instead to the mismatch between data-driven machine learning methods and the demands of human interpretation (Burrell, 2016). Since machine learning methods are designed to discover subtle correlations, and not to understand the causes behind any phenomenon being studied, the resulting classifiers often take the form of complicated formulae that are not explainable to the general user. While it is true that these formulae can be proven to be predictive and, under certain assumptions, even optimal, it is not possible for an individual decision to be explained. A random forest, a boosted-combination of trees, or even a linear combination, cannot provide a unvarnished explanation for a citizen whose liberty has been denied by the algorithm.

Returning to the HART case study, it would be unfeasible for a human being to look for an explanation in nearly 4.5 million nested and conditionally-dependent decision points spread across many hundreds of unique trees (Oswald *et al.* 2018). Paraphrasing Burrell (2016), we might say that when HART learns and builds its own representation of a classification decision, it does so without regard for human comprehension. So, even if a learnt representation was available it would be extremely difficult, even for an expert, to fully understand and interpret the decision process (e.g. to understand the impact of a specific feature).

In recent years the research community has developed several ways to characterise and improve the transparency of machine decisions (Lipton 2016). Some of these amount to post-hoc interpretations, i.e. separate models which furnish verbal justifications or more simple representation of algorithmic output (e.g. visualisation of active nodes in a neural network). Other approaches try to justify a decision by reporting analogous cases (i.e. a model generating examples that are similar). An alternative strategy is to build systems that ensure *procedural regularity,* by verifying that an automated decision meets specific policy goals through software verification and cryptographic techniques (Kroll *et al.* 2017).

As well as technical remedies, transparency might be supported by sectoral laws, such as the US Fair Credit Reporting Act 1970 (through the creation and maintenance of a consumer report), or by laws of general application such as the EU GDPR, which acknowledges the right of the data subject to access personal data which have been collected concerning them (Article 15), and requires data controllers to implement appropriate measures to safeguard the data subject's rights and freedoms and legitimate interests, including the right to obtain human intervention on the part of the controller (Article 22 and Recital 71).

**Informational privacy and freedom of expression**

The type of input data in risk assessment tools like HART largely reflect information that police officers collect about inmates at specific steps of the justice process (e.g. at intake), such as a suspect's demographics and criminal history.[47] This data collection is mandated and/or supported by legislative provisions, but the ability to integrate heterogeneous sources of information and other social factors (e.g. greater levels of perceived vulnerability among citizens) might suggest further directions. For example, in New Orleans a software program (owed by Palantir) is in use by the police department to identify connections between individuals and gang members by analysing both criminal histories and social media activity (Winston 2018). This has implications for both privacy and freedom of expression – if individuals are aware of the fact that their behaviour or interactions with others may be used to make predictions about them and that consequential decisions may be made about them as a result of those predictions, they may be caused to consciously or subconsciously alter those behaviours and interactions. In the case of the New Orleans example, this might involve possible targets not using social media, or changing the style or content of their communication with the consequence of 'silencing' them or restricting their ability to communicate effectively.

---

[47] The business of risk assessment tools is associated with the development of information management systems (sometimes the former is considered an extension of the latter, see e.g. CourtView Justice Solutions: http://www.courtview.com/court)

The use of personal data in machine decision-making raises serious privacy concerns: what data are admissible? In which contexts? Should public expressions of a citizen be permitted to be used as part of computing his risk or credit score? More generally, can personal private online activities be used as an element in assessing the score of individuals? Might this directly or indirectly undermine their right to expression?

The right to (informational) privacy is commonly reframed as the "right to control over personal information"[48] which refers to each piece of data that can be linked to a person ("data subject") such as date of birth, social security number, fingerprint, etc.[49] Note that when personal data refers to intimate aspects of life or valuable information such as ethnicity, political opinions, religious or philosophical beliefs, biometric data, sexual orientation, etc. is also called "sensitive data".

From a technical standpoint, the field of computer science has been developing privacy enhancing technologies for many years, for instance, introducing privacy concerns in data release and processing (e.g. k-anonymity and differential privacy) or addressing the problem of data disclosure in social networking platforms. All these efforts have produced an exceptional stock of tools for those who would be willing to take up privacy by design[50] and, considering the growing concerns about information privacy and security, we would expect further progress on this camp.

The simplest method to tackle privacy is to de-identify data, i.e. to remove, generalize or replace with made-up alternatives explicit identifiers such as name, social security number, address, phone number, etc. (Sweeney, 1997). Unfortunately, deidentification is often a limited solution since the combination of de-identified databases with other publicly available data ("auxiliary information") can reveal subjects' identity or reduce uncertainty on a subset of subjects in the dataset.[51]

Another technical approach is provided by the model of k-anonymity (Sweeney 2002), a property that characterizes the degree of privacy protection in the release of a database. Roughly speaking, k-anonymity requires that every combination of the values reported in the data released can be indistinctly matched to at least k individuals in the database. However, k-anonymity also presents some weaknesses as it assumes restrictions on auxiliary information (i.e. an attacker's knowledge may be greater than what is supposed to be) and does not perform well in high-dimensional datasets (Narayanan & Shmatikov 2008).

Contrary to traditional models, the framework of differential privacy (Dwork 2006) shifts the focus from database to computation by drawing on randomized response in the structured interview (a technique that allows study participants to answer embarrassing questions while maintaining confidentiality). Differential privacy aims at the same time to maximize the accuracy of data analysis and minimize the risk of re-identification. Given a database and a series of queries, differential privacy ensures that "that any sequence of outputs (responses to queries) is "essentially" equally likely to occur, independent of the presence or absence of any individual" in the database (Dwork 2014). Note that while the theoretical research has rapidly grown, the concrete application of differential privacy started more recently.[52]

---

[48] For a different account of privacy, see Nissenbaum (2010)

[49] In the EU legal terms personal data indicates "any information relating to an identified or identifiable natural person ('data subject')" where "an identifiable natural person is one who can be identified, directly or indirectly, in particular by reference to an identifier such as a name, an identification number, location data, an online identifier or to one or more factors specific to the physical, physiological, genetic, mental, economic, cultural or social identity of that natural person" (GDPR ).

[50] Privacy by design is a methodological approach that supports the inclusion of privacy into the whole lifecycle of IT systems (Cavoukian 2009).

[51] The release of AOL web search data (Hansell 2006) and the Netflix competition (Narayanan & Shmatikov 2008) are two popular examples of failed anonymisation.

[52] See, e.g. Apple's announcement (Greenberg 2016).

**CONCLUDING REMARKS**

Machine decision-making is set to become an increasing factor in our lives. The calculation whether to adopt such technology may be driven by considerations of promised accuracy and efficiency gains, or simply by the attraction of downsizing and cost-savings. Yet the social impacts of removing or reducing human interventions in decision-making that will impact individual rights and personal opportunities, as well as the collective good, will be wide-ranging. To date, the speed of adoption of such technologies in sensitive administrative and criminal decision-making processes has clearly outpaced evaluation of, and reaction to, those impacts. Contemporary research raises issues as diverse as the deskilling and devaluation of certain types of employment, e.g. social workers become call centre workers (Eubanks 2018); loss of social connection, critical judgement and empathy as decisions become divorced from personal interactions (Turkle 2015); and social exclusion of the poorly educated or those in poverty who lack the resources to identify, understand and challenge system errors and failings (Eubanks 2018).

This chapter addresses one of these impacts: where algorithms make decisions that seek to evaluate individual persons for the purposes of making a decision about their entitlement to some tangible benefit or burden, it is essential that those decisions and the system within which they are made remain capable of incorporating and acting upon appropriate normative principles and obligations. While machine decisions are often judged based solely on their accuracy, in such circumstances there are wider normative considerations. Thus, the use of machine learning-based decision-making in criminal justice decisions about bringing criminal charges, length of prison sentence, eligibility for parole; or state administrative decisions about child safeguarding and eligibility for welfare benefits differs fundamentally from its use in other spheres, even those that may also have a positive or negative impact on individuals, such as medical diagnosis decisions.[53]

This chapter reviewed some technical aspects of machine decisions and suggests that when seeking to develop or maintain a trust framework that supports the social legitimacy of criminal justice or state administrative decision-making systems using machine decision-making, it is helpful to do so by reference to key representative benchmarks. Such benchmarks are common to consequential decision-making relating to individuals, and critical to development of appropriate trust relations between stakeholders, and include accuracy, transparency/accountability, fairness/equal treatment, and rights to privacy/free expression. They reflect the main tracks around which research communities have articulated the main social and ethical implications of learning algorithms in the big data context (e.g. Fairness, Accountability, and Transparency in Machine Learning (FAT/ML) scholarship). In a similar vein, Mittelstadt *et al.* (2016) have framed the discourse with a distinctive vocabulary highlighting both epistemic and normative concerns.

Each of these benchmarks may incorporate a multitude of interrelated aspects: legal concepts, metrics, (technical and legal) procedures, values, rights, existing laws, pragmatic concerns, etc. In this paper we have tried to demonstrate some of the connections between these aspects to shed light on the variety of (human) choices involved in the design, assessment and deployment of such systems (e.g. dealing with fairness requires consideration for different notions of equality and, in some contexts, existing laws, understanding of distinct technical criteria and techniques, etc.). Choosing which set of notions or procedures to follow calls for in-depth analysis, assessment of commitments and delicate balance. Of course, the benchmarks identified here do not exist independently of one another. Rather they are in constant interaction and many problems arise out of their fuzzy boundaries, consider the use of sensitive data in decision making and the risk of discrimination. This also has consequences for the search for, and the implementation of, remedies for particular problems, e.g. an intervention on data protection/privacy may impact negatively on concerns regarding fairness.

It is recognised that while 'trust' is frequently referenced when members of a society are affected by the deployment of novel infrastructures, such as socio-technical systems, the precise meaning and nature of trust has been debated and contested at length in the social sciences literature. We appreciate that the brief discussion here of the relationship of trust/distrust to the legitimacy of criminal justice or state administrative decision-making systems, and to the utility of the suggested benchmarks, cannot begin to do justice to that literature. Nevertheless, we would argue that the inclusion of the normative principles and obligations reflected in the benchmarks is critical to the social acceptance of those systems. The principles and obligations provide rational grounds to engage with such decision-making systems and to accept their decisions: providing a reason to trust. They also facilitate meaningful criticism and oversight of decision-making processes within those systems: permitting the ability to distrust.

---

[53] As opposed to medical treatment decisions – whether a person has a particular form of cancer is a different question to whether the person should receive treatment for the cancer, and if so, what that treatment should be.

Most of the social and ethical issues raised by these benchmarks have also been tackled by Oswald *et al.* (2018), who developed a framework (ALGO-CARE) drawing on the experience of Durham Constabulary. The framework translates key ethical and legal principles in considerations and questions for supporting the assessment of predictive tools like HART. We hope that this paper can contribute to the discussion and the development of guidelines like ALGO-CARE with a synthesis of relevant technical and legal notions. A purely algorithmic solution to these problems is not likely to be sufficient, as many of these problems require the development of legal and cultural tools and available metrics cannot capture the entire spectrum of human values and principles. Our ambition was to disclose important parts of the process implied in assessing our trust in machine decision-making, and to make accessible some conceptual tools that are necessary to approach the issues raised by machine learning algorithms.

**Acknowledgments.** Teresa Scantamburlo and Nello Cristianini were funded by the ERC Advanced Grant ThinkBIG.

Zafar MB, Valera I, Rodriguez MG, & Gummadi KP (2017). Fairness beyond disparate treatment & disparate impact: Learning classification without disparate mistreatment. *Proceedings of the 26th International Conference on World Wide Web:* 1171-1180. *arXiv*:1610.08452

Zliobate I (2015) A survey on measuring indirect discrimination in machine learning. *arXiv*:1511.0014826